\documentclass[aps,prd,preprintnumbers,nofootinbib,twocolumn]{revtex4}
\usepackage{bm}
\usepackage{latexsym}
\usepackage{dcolumn}
\usepackage{amsmath,amsfonts,amssymb}
\usepackage{graphicx,epsfig}
\usepackage{psfrag}
\usepackage{amsthm}

\def\be {\begin{equation}}
\def\ee {\end{equation}}
\def\bea {\begin{eqnarray}}
\def\eea {\end{eqnarray}}
\def\bc {\begin{center}}
\def\ec {\end{center}}
\def\bfg {\begin{figure}}
\def\efg {\end{figure}}
\def\bi {\begin{itemize}}
\def\ei {\end{itemize}}
\def\nn {\nonumber}

\def\la {\label}
\def\le {\left}
\def\ri {\right}

\def\fr {\frac}

\def\no {\noindent}

\def\vs {\vspace}

\def\ul {\underline}

\def\b  {\beta}

\def\d  {\delta}

\def\beq{\begin{equation}}
\def\eeq{\end{equation}}
\def\br{\begin{eqnarray}}
\def\er{\end{eqnarray}}
\newcommand{\eel}[1] {\label{#1}\end{equation}}

\newcommand{\bdm}{\begin{displaymath}}
\newcommand{\edm}{\end{displaymath}}

\begin{document}
\title{Universality of Quantum Gravity Corrections}


%

\author{Saurya Das}
\email{ saurya.das@uleth.ca}

\affiliation{Department of Physics, University of Lethbridge,\\
 4401 University Drive, Lethbridge, Alberta -- T1K 3M4, Canada}

\author{Elias C. Vagenas}
\email{evagenas@academyofathens.gr}

\affiliation{Research Center for Astronomy \& Applied Mathematics,\\
Academy of Athens, \\
Soranou Efessiou 4,
GR-11527, Athens, Greece}

\begin{abstract}
We show that the existence of a minimum measurable length and the
related Generalized Uncertainty Principle (GUP), predicted by
theories of Quantum Gravity, influence all quantum Hamiltonians.
Thus, they predict quantum gravity corrections to various quantum
phenomena. We compute such corrections to the Lamb Shift, the
Landau levels and the tunnelling current in a Scanning Tunnelling
Microscope (STM). We show that these corrections can be
interpreted in two ways: (a) either that they are exceedingly
small, beyond the reach of current experiments, or (b) that they
predict upper bounds on the quantum gravity parameter in the GUP,
compatible with experiments at the electroweak scale. Thus, more
accurate measurements in the future should either be able to test
these predictions, or further tighten the above bounds and predict
an intermediate length scale, between the electroweak and the
Planck scale.
%
\end{abstract}

\maketitle


We know that gravity is universal. Anything which has
energy creates gravity and is affected by it, although the smallness of Newton's
constant $G$ often means that the associated effects are too weak to be measurable.
In this article, we show that certain effects of Quantum Gravity are also universal,
and can influence almost any system with a well-defined Hamiltonian.
The resultant effects are generically quite small,
being proportional to the
square of the Planck length $\ell_{Pl}^2=G\hbar/c^3$. However, with
current and future experiments,
bounds may be set on certain parameters relevant to quantum gravity,
and improved accuracies could even make them measurable.

An important prediction of various theories of quantum gravity
(such as String Theory) and black hole physics is
the existence of a minimum measurable length.
The prediction is largely model
independent, and can be understood as follows: the Heisenberg
Uncertainty Principle (HUP), $\Delta x \sim \hbar/\Delta p$, breaks down
for energies close to the Planck scale, when the corresponding
Schwarzschild radius is comparable to the Compton wavelength
(both being approximately equal to the Planck length).
Higher energies result in a further increase of
the Schwarzschild radius, resulting in $\Delta x \approx \ell_{Pl}^2\Delta p/\hbar$.
The above observation, along with a combination of thought experiments and
rigorous derivations suggest that the {\it Generalized Uncertainty
Principle} (GUP) holds at all scales, and is represented by
\cite{guppapers}
%
\bea \Delta x_i \Delta p_i &\geq& \fr{\hbar}{2} [ 1 + \beta
\le((\Delta p)^2 + <p>^2 \ri) \nn \\
&+& 2\beta \le( \Delta p_i^2 + <p_i>^2\ri) ]~,~i=1,2,3 \la{uncert1}
\eea
where $p^2 = \sum\limits_{j=1}^{3}p_{j}p_{j}$,
$\beta=\beta_0/(M_{Pl}c)^2=\ell_{Pl}^2/2\hbar^2$, $M_{Pl}=$
Planck mass, and $M_{Pl} c^2=$ Planck energy $\approx 1.2 \times
10^{19}~GeV$.
Implications of the GUP in various fields, including High Energy Physics,
Cosmology and Black Holes, have been studied.
Here, we examine its potential experimental
signatures in some familiar quantum systems.
It is normally assumed that the dimensionless parameter
$\beta_0$ is of the order of unity. However, as we shall see in this
article, this choice renders Quantum Gravity effects too small to be measurable.
%
%
%
On the other hand, if one does not impose the above condition {\it a
priori}, current experiments predict large upper bounds on it,
which are compatible with current observations, and may signal the
existence of a new length scale. Note that such an intermediate length
scale, $\ell_{inter} \equiv \sqrt{\beta_0} \ell_{Pl}$ cannot
exceed the electroweak length scale $\sim 10^{17}~\ell_{Pl}$ (as
otherwise it would have been observed). This implies $\b_0 \leq 10^{34}$.
(The factor of $2$ in the last term
in Eq.(\ref{uncert1}) follows from Eq.(\ref{com1}) below).

It was shown in \cite{kmm}, that inequality (\ref{uncert1}) is equivalent to the
following modified Heisenberg algebra
\be [x_i,p_j] = i \hbar ( \delta_{ij} + \beta \delta_{ij} p^2 +
2\beta p_i p_j )~. \la{com1} \ee
This form ensures, via the Jacobi identity, that
$[x_i,x_j]=0=[p_i,p_j]$ \cite{kempf}.
Next, defining
\bea x_i = x_{0i}~,~~
p_i = p_{0i} \le( 1 + {\beta} p_0^2 \ri) \la{mom1}
\eea
where $p_0^2 = \sum\limits_{j=1}^{3}p_{0j}p_{0j}$ and $x_{0i}, p_{0j}$
satisfying the canonical commutation relations
%
$ [x_{0i}, p_{0j}] = i \hbar~\delta_{ij}, $
%
it is easy to show that Eq.(\ref{com1}) is satisfied
to order $\beta$ (henceforth we neglect terms of order
$\beta^2$ and higher). Here,
$p_{0i}$ can be interpreted as the momentum at low energies
(having the standard representation in position space, i.e. $p_{0i} = -i
\hbar d/d{x_i}$), and $p_{i}$ as that at higher energies.

Using (\ref{mom1}), any Hamiltonian of the form
\bea H &=& \fr{p^2}{2m} + V(\vec r)
~~~[\vec r =(x_1,x_3,x_3)]
\eea
can be written as \cite{brau}
\bea H&=&
H_0 + H_1 + {\cal O}( \beta^2) ~, \\
%
%
\mbox{where}~~H_0 &=& \fr{p_0^2}{2m} + V(\vec r) 
~~\mbox{and}~~ H_1 = {\fr{\beta}{m} p_0^4} ~\la{H0}.\eea
Thus, we see that {\it any} system with a
well defined quantum (or even classical) Hamiltonian $H_0$, is perturbed
by $H_1$, defined above, near the Planck scale. In other words,
Quantum Gravity effects are in some sense universal! It remains to
compute the corrections to various phenomena due to the
Hamiltonian $H_1$.
%
%
In this article, we study its effects on three such
well-understood quantum phenomena, the {\bf Lamb shift}, the {\bf Landau
levels} and the {\bf Scanning
Tunnelling Microscope}.

\vs{0.2cm} \no \ul{\bf I. The Lamb shift}
\vs{.1cm}

For the Hydrogen atom, $V(\vec r) = -k/r$ ($k=e^2/4\pi
\epsilon_0=\alpha \hbar c$,
$e=$ electronic charge). 
To first order, the perturbing Hamiltonian
$H_1$, shifts the wave-functions to \cite{bransden}
\bea |\psi_{nlm} \rangle_1 = |\psi_{nlm} \rangle +\hspace{-3ex}
\sum_{\{n'l'm'\} \neq \{ nlm\}} \hspace{-3ex}\frac{e_{n'l'm'|nlm}}{E_{n} -
E_{n'}} |n'l'm'\rangle \la{wavefn1}
\eea
where $n,l,m$ have their usual significance, and
$e_{n'l'm'|nlm} \equiv \langle n'l'm'|H_1|nlm\rangle$~.
Using $p_0^2=2m[H_0 + k/r]$ \cite{brau}
\be H_1 = (4\beta m) \le[ H_0^2 + k \le( \frac{1}{r}H_0 + H_0
\frac{1}{r} \ri) + \le( \frac{k}{r}\ri)^2 \ri] ~.\ee
\par\noindent
Thus,\vspace{-4ex}\\
\bea &&
\frac{ e_{n'l'm'|nlm}}{4\beta m} = \le( E_n \ri)^{2}\delta _{n n'} \nn \\
&& + k \le( E_n + E_{n'} \ri)
\langle n'l'm'|\frac{1}{r} | nlm \rangle 
+ k^2 \langle n'l'm'|\frac{1}{r^2} | nlm \rangle~. \la{mat0} \nn
\eea
It follows from the orthogonality of spherical harmonics that the
above are non-vanishing if and only if $l'=l$ and $m'=m$. Thus,
the first order shift in the ground state wave-function is given
by (in the position representation)\vspace{-3ex}\\
\bea
\Delta \psi_{100}(\vec r) &\equiv& \psi_{100(1)}(\vec
r)-\psi_{100}(\vec r) =
\frac{e_{200|100}}{E_1-E_2}\psi_{200} (\vec r)\nn\\
&=& \frac{928\sqrt{2} \beta m E_0}{81}~\psi_{200}(\vec r) ~. \eea
%
\vspace{-6ex}\\
\par\noindent
Next, consider the Lamb shift for the $n^{th}$
level of the Hydrogen atom
\cite{bd}
\be
\Delta E_n = \frac{4\alpha^2}{3m^2} \le( \ln \frac{1}{\alpha} \ri) \le|
\psi_{nlm}(0) \ri|^2~.
\ee
Varying $\psi_{nlm}(0)$, the additional contribution
due to GUP in proportion to its original value is given by
\be
\frac{\Delta E_{n(GUP)}}{\Delta E_n} =
2 \frac{\Delta|\psi_{nlm}(0)|}
 {\psi_{nlm}(0)}~.
\ee
Thus, for the Ground State, using $\psi_{100}(0) =
a_0^{-3/2}\pi^{-1/2}$ and $\psi_{200}(0)=
a_0^{-3/2}(8\pi)^{-1/2}$, where $a_0$ is the Bohr radius, we get
\bea \frac{\Delta E_{0(GUP)}}{\Delta E_0} &=& 2
\frac{\Delta|\psi_{100}(0)|}
 {\psi_{100}(0)}= \frac{928 \beta m E_0}{81} \nn\\
 &\approx&\hspace{-1ex} 10 \b_0 \frac{m}{M_{Pl}}
 \frac{E_0}{M_{Pl} c^2}
%
 \approx 0.47 \times 10^{-48}\hspace{-1ex}~\b_0.
\eea
The above result may be interpreted in two ways. First, if one assumes
$\beta_0 \sim 1$, then it predicts a non-zero, but virtually
{\it unmeasurable} effect of Quantum Gravity/GUP. On the other hand, if such
an assumption is not made, the current accuracy of
precision measurement of Lamb shift of about
$1$ part in $10^{12}$ \cite{brau,newton}, sets the following upper bound on $\b_0$
\vspace{-3ex}\\
\be
\b_0 < 10^{36}~.
\la{beta1}
\ee
This bound is weaker than that set by the electroweak scale, but not
incompatible with it.
Moreover, with more accurate measurements in the future, this bound is expected to get
reduced by several orders of magnitude, in which case, it could signal a new and
intermediate length scale between the electroweak and the Planck scale.

\vs{0.2cm} \no \ul{\bf II. The Landau Levels}
\vs{0.1cm}

Next consider a particle of mass $m$ and charge $e$ in a constant
magnetic field ${\vec B} = B {\hat z}$, described by the vector
potential ${\vec A}=Bx {\hat y}$ 
and the Hamiltonian
\bea H_0 &=& \frac{1}{2m}\le( \vec p_0 - e \vec A\ri)^2  \la{lanham1}\\
&=& \frac{p_{0x}^2}{2m} + \frac{p_{0y}^2}{2m} - \frac{eB}{m}~x p_{0y} +
\frac{e^2 B^2}{2m}~x^2~. \la{lanham2}
\eea
Since $p_{0y}$ commutes with $H$, replacing it with its eigenvalue
$\hbar k$, we get
\be H_0 = \frac{p_{0x}^2}{2m} + \frac{1}{2} m \omega_c^2 \le( x -
\frac{\hbar k}{m \omega_c}\ri)^2 ~,\la{lanham4}\ee
where $\omega_c=eB/m$ is the the cyclotron frequency. This is nothing
but the Hamiltonian of a harmonic oscillator in $x$ direction,
with its equilibrium position given by $x_0 \equiv \hbar k/m
\omega_c$. Consequently, the eigenfunctions and eigenvalues are
given by
\bea \psi_{k,n} (x,y) &=& e^{ik y} \phi_n (x-x_0) \\
E_n &=& \hbar \omega_c \le( n +\frac{1}{2} \ri)~,~n\in N~, \eea
where $\phi_n$ are the harmonic oscillator wave-functions.

Following the procedure outlined in the introduction, the
GUP-corrected Hamiltonian assumes the form
\bea H &=& \frac{1}{2m}\le( \vec p_0 - e \vec A\ri)^2 +
\frac{\beta}{m}\le( \vec p_0 - e \vec A\ri)^4
\la{lanham13} \\
&=& H_0 + 4\b m H_0^2
%
\eea
where in the last step we have used Eq.(\ref{lanham1}).
Evidently, the eigenfunctions remain unchanged, which alone guarantees for
example, that the GUP will have no effect at all on phenomena such
as the Quantum Hall Effect, the Bohm-Aharonov effect, and Dirac
Quantization. However, the eigenvalues shift by
\bea
\Delta E_{n(GUP)}\hspace{-1ex} &=&\hspace{-1ex} 4\b m \langle \phi_n |H_0^2| \phi_n \rangle 
\hspace{-0.5ex}=\hspace{-0.5ex} 4\b m (\hbar \omega_c )^2\hspace{-1ex} \le( n + \frac{1}{2} \ri)^2 \hspace{-1ex}, \nn \\
%
%
\mbox{or}~~
\frac{\Delta E_{n(GUP)}}{E_n}\hspace{-1ex} &=&\hspace{-1ex} 4\b m (\hbar \omega_c ) \le( n + \frac{1}{2} \ri) 
\hspace{-0.5ex}\approx\hspace{-0.5ex} \beta_0 \frac{m}{M_{Pl}} \frac{\hbar \omega_c}{M_{Pl}c^2}~. \nn
%
\eea
%
%
\par\noindent
For an electron in a magnetic field of $10~T$, $\omega_c \approx 10^3~GHz$ and we get
%
%
\bea
\frac{\Delta E_{n(GUP)}}{E_n}
&\approx&
2.30 \times 10^{-54} \b_0~.
\eea
%
%
Once again, if $\beta_0 \sim 1$, this correction is too small
to be measured. Without this assumption,
an accuracy of $1$ part in $10^3$ in direct measurements
of Landau levels using a STM (which is somewhat
over-optimistic) \cite{wildoer}, the upper bound on $\b_0$ follows
\be
\b_0 < 10^{50}~.
\la{beta2}
\ee
This bound is far weaker than that set by electroweak
measurements, but compatible with the latter (as was the case for
the Lamb shift). Once again, better accuracy should tighten this
bound, and perhaps predict an intermediate length scale.

\vs{0.2cm} \no \ul{\bf III. Potential Barrier and STM}
\vs{0.1cm}

In a STM, electrons of energy $E$ (close to the Fermi energy)
from a metal tip at $x=0$, tunnel quantum mechanically to
a sample surface a small distance away at $x=a$. This gap,
(across which a bias voltage may be applied), is associated with
a potential barrier of height $V_0>E$ \cite{stroscio}.
Thus
\be
V(x) = V_0~\le[ \theta(x) - \theta(x-a) \ri]~,
\ee
where $\theta(x)$ is the usual step function.
The wave-functions for the
three regions, namely $x\leq 0$, $0\leq x \leq a$, and $x\geq a$, are $\psi_{1}$,$\psi_{2}$,
and $\psi_{3}$, respectively, and satisfy the following
GUP corrected Schr\"odinger equation ($d^n \equiv d^n/dx^n$)
\bea
&& d^2\psi_{1,3} + k^2 \psi_{1,3} - \ell_{Pl}^2 d^4 \psi_{1,3} = 0~,\nn \\
&& d^2\psi_2 -k_1^2 \psi_2 - \ell_{Pl}^2 d^4 \psi_2 = 0~, \nn
%
%
\eea
where
$ k=\sqrt{2mE/\hbar^2}~,~k_1=\sqrt{2m(V_0-E)/\hbar^2}~.$
%
%
%
%
The solutions to the above to leading order in $\ell_{Pl}, \b$ are
\bea
\psi_1 &=& Ae^{ik'x} + B e^{-ik'x} + A_1 e^{x/\ell_{Pl}}~~, \la{barrierpsi1} \\
\psi_2 &=& Fe^{k_1'x} + Ge^{-k_1'x} + H_1e^{x/\ell_{Pl}} + L_1
e^{-x/\ell_{Pl}},
\la{barrierpsi2} \\
\psi_{3} &=& Ce^{ik'x} + D_1 e^{-x/\ell_{Pl}}~~, \la{barrierpsi3}
\eea
where $k'\equiv k(1-\beta \hbar^2 k^2 )$ and $k_1'\equiv k_1(1-\beta \hbar^2 k_1^2 )$.
Note the appearance of new exponential terms,
which drop out in the $\ell_{Pl} \rightarrow 0$ limit.
In the above, we have omitted the left-mover from $\psi_{3}$ and the exponentially growing
terms from both $\psi_1$ and $\psi_{3}$.
The boundary conditions
\bea
d^n\psi_1|_{x=0} &=& d^n\psi_2|_{x=0}~~,~n=0,1,2,3~ \\
d^n\psi_2|_{x=a} &=& d^n\psi_{3}|_{x=a}~~,~n=0,1,2,3~
\eea
on wave-functions (\ref{barrierpsi1}-\ref{barrierpsi3})
%
%
%
yield the following transmission coefficient
\bea
%
%
T &=& \le| \frac{C}{A} \ri|^2
= \le[
1 + \frac{(k'^2 + k_1'^2)^2\sinh^2(k_1'a)}{(2k'k_1')^2}
\ri]^{-1}~. \la{trans3}
\eea
The reflection coefficient $R=|B/A|^2 =1-T$.
Using Eq.(\ref{trans3}) and the definitions of $k,k_1,k',k_1'$,
it can be shown that when $k_1 a \gg 1$, which is the limit relevant for
STMs, the transmission coefficient is approximately
\bea
 \hspace{-1ex}T\hspace{-1ex}&=&\hspace{-1ex}T_0\hspace{-0.5ex}\le[
1\hspace{-0.5ex} +\hspace{-0.5ex} \frac{4m\b (2E-V_0)^2}{V_0}\hspace{-0.5ex}+\hspace{-0.5ex}\frac{2\b a}{\hbar} [2m (V_0-E)]^{\frac{3}{2}}
\ri] \la{stmt} \\
&&~~~~ {\rm where}~~T_0 =
\frac{16E(V_0-E)}{V_0^2} e^{-2k_1a}~,
\eea
$T_0$ being the standard tunnelling amplitude.
The current $I$ flowing from the tip to the sample
is proportional to $T$, and
is usually magnified using an amplifier of gain ${\cal G}$.
From Eq.(\ref{stmt}) the enhancement in current due to GUP is given by
\bea
&& \frac{\d I}{I_0} = \frac{\d T}{T_0}
= 4\b \frac{m (2E-V_0)^2}{V_0} + \frac{2\b a}{\hbar} [2m (V_0-E)]^{\frac{3}{2}} \nn \\
&=& \frac{4\b_0 m}{M_{Pl}} \frac{(2E-V_0)^2}{V_0 M_{Pl} c^2 } 
+
\frac{ 4\sqrt{2} {\b_0} a}{\ell_{Pl}}
\left( \frac{m}{M_{Pl}} \ri)^{\frac{3}{2}}\hspace{-0.5ex}
\left( \frac{V_0-E}{M_{Pl}c^2} \ri)^{\frac{3}{2}}. \nn
\eea
Then, assuming the following approximate (but realistic) values \cite{stroscio}
\bea
&& m = m_e = 0.5~MeV/c^2 ~,~
E\approx V_0 = 10~eV~,  \nn \\ 
&& a = 10^{-10} ~m ~,~
I_0 = 10^{-9} ~A~,~
{\cal G} = 10^9~, \nn
\eea
%
%
%
%
%
we get
\be
\frac{\d I}{I_0} = \frac{\d T}{T_0} = 10^{-48} \b_0
~\mbox{and}~ \d {\cal I} \equiv {\cal G} \d I = 10^{-48} \b_0~A ~.
\ee
Thus, for the GUP induced
excess current $\d{\cal I}$ to add up to the charge of just one electron,
$e \approx 10^{-19}~C$, one would have to wait for a time
\be
\tau = \frac{e}{\d{\cal I}}  = 10^{29} \b_0^{-1}~s~.
\ee
If $\b_0 \sim 1$, this is far greater than the age of our universe ($10^{18}~s$).
However, if the quantity $\d {\cal I}$ can be increased
by a factor of about $10^{21}$, say by a combination of increase in
$I$ and ${\cal G}$, and by a larger value of $\b_0$, the above time will be reduced
to about a year ($\approx 10^8~s$), and one can hope to measure the effect of GUP.
Conversely, if such a GUP induced current cannot be
measured in such a time-scale, it will put an upper bound
\be
\b_0 < 10^{21}~. \la{beta3}
\ee
Note that this is more stringent than the two previous examples, and
is in fact consistent with that set by the electroweak scale!
In practice however, it may be easier to experimentally determine the
{\it apparent barrier height} $\Phi_A \equiv V_0-E$,
and the (logarithmic) rate of increase of current with the gap. From Eq.(\ref{stmt})
they are related by \cite{stroscio}
\be
\sqrt{\Phi_A} = \frac{\hbar}{\sqrt{8m}}
\le|\frac{d\ln I}{da}\ri| \le( 1 - \frac{\beta \hbar^2}{4}
\le| \frac{d\ln I}{da}\ri|^2 \ri)~.
\ee
\par\noindent
The cubic deviation from the linear $\sqrt{\Phi_A}$ vs
$\le| \frac{d\ln I}{da} \ri|$ curve predicted by GUP may be easier to spot
and the value of $\b$ estimated with improved accuracies.

\par
To summarize, our results
indicate that either the predictions of GUP are too small
to measure at present ($\beta_0 \sim 1$), or that they signal
a new intermediate length scale ($\beta_0 \gg 1$).
It is not inconceivable that such a new length scale may
show up in future experiments in the Large Hadron Collider.
%
%
Perhaps more importantly, our study reveals the universality of GUP effects,
meaning that the latter can potentially be tested in a wide class of quantum mechanical
systems, of which we have studied just a handful here.
Promising areas include statistical systems (where a large number of particles
may offset the smallness of $\beta$), study of whether normally forbidden
transitions and processes can be allowed by the GUP corrected
Hamiltonian, and processes which may get corrected by a fractional power
of $\beta_0$. We hope to report on some of these in the near future.
In the best case scenario, this could open a much needed low-energy
`window' to Quantum Gravity Phenomenology.
\vspace{-2ex}\\


\no {\bf Acknowledgment}

We thank K. Ali, B. Belchev, A.
Dasgupta, R. B. Mann, S. Sur and M. Walton for useful
discussions. This work was supported in part by the Natural
Sciences and Engineering Research Council of Canada and by the
Perimeter Institute for Theoretical Physics.



\end{document}